\documentclass[]{spie}  

\usepackage{amsmath,amsfonts,amssymb}
\usepackage{graphicx}
\usepackage{subfigure}
\usepackage[colorlinks=true, allcolors=blue]{hyperref}
\usepackage{svg}
\usepackage{gensymb}
\usepackage{amssymb}
\usepackage{wasysym}
\usepackage{textcomp}
\usepackage{tcolorbox}

\title{Atmospheric dispersion corrector for a multi-object spectroscopic mode of HROS-TMT}

\author[a,b]{Manjunath Bestha}
\author[a]{Amirul Hasan}
\author[a]{Devika Divakar}
\author[b]{Arun Surya}
\author[a]{S. Sriram}
\author[a]{T. Sivarani}
\author[a]{Ajin Prakash}
\author[a,b]{Parvathy M}
\author[a]{Sudharsan Yadav}
\affil[a]{Indian Institute of Astrophysics, Bangalore, India}
\affil[b]{University of Calcutta, India}
\affil[c]{Tata Institute of Fundamental Research, Mumbai, India}

\authorinfo{Further author information(Send correspondence to Manjunath Bestha): \\ Manjunath Bestha: E-mail: manjunath.b@iiap.res.in\\  Sivarani Thirupathi: E-mail: sivarani@iiap.res.in}

\begin{document} 

\maketitle

\begin{abstract}
Highly multiplexed spectroscopic surveys have changed the astronomy landscape in recent years. However, these surveys are limited to low and medium spectral resolution. High spectral resolution spectroscopy is often photon starved and will benefit from a large telescope aperture. Multiplexed high-resolution surveys require a wide field of view and a large aperture for a suitable large number of bright targets. This requirement introduces several practical difficulties, especially for large telescopes, such as the future ELTs. Some of the challenges are the need for a wide field atmospheric dispersion corrector and to deal with the curved non-telecentric focal plane. Here we present a concept of Multi-Object Spectroscopy (MOS) mode for TMT High-Resolution Optical Spectrograph (HROS), where we have designed an atmospheric dispersion corrector for individual objects that fit inside a fiber positioner. We present the ZEMAX design and the performance of the atmospheric dispersion corrector for all elevations accessible by TMT.

\end{abstract}


\section{INTRODUCTION}
\label{sec:intro}  

Ground-based telescopes are affected due to the wavelength-dependent refraction of light as it passes through the Earth's atmosphere. Due to the varying refractive index of air with wavelength, different wavelengths are refracted to varying degrees. Large plate scales make this dispersion effect appears particularly prominent for large telescopes. With increasing telescope apertures and advanced instruments, precise correction techniques have become crucial to minimize image degradation and enable high-quality observations across a wide range of wavelengths. However, when it comes to large telescopes like the upcoming Extremely Large Telescope (ELT) and Thirty Meter Telescope (TMT), correcting dispersion for an entire wide field of view is challenging because it would require large correctors. In this paper, we present the design of an Atmospheric Dispersion Corrector (ADC) that focuses explicitly on the correction of multiple single objects that corresponds to individual narrow fields. The designed ADC is intended to feed dispersion-corrected light to the High-Resolution Optical Spectrograph (HROS), a second-generation instrument proposed for the Thirty Meter Telescope (TMT). HROS covers a broad wavelength range from 310 to 1000nm and operates in various modes, including a possible Multi-Object Spectroscopy mode (MOS). In MOS mode, six objects can be observed simultaneously covering the entire wavelength range, each utilizing a 1" fiber to achieve a resolution of 25000. Blocking subsequent echelle orders, up to forty objects could be accommodated in the HROS slit. The light from the focal plane is channeled to the spectrograph using fiber positioners \cite{Sivarani_2022}, within which the atmospheric dispersion corrector can be placed \cite{zebri_2014}. Two primary types of atmospheric dispersion correctors are commonly used, Linear Atmospheric Dispersion Corrector (LADC) and Rotational Atmospheric Dispersion Corrector (RADC). The LADC operates in a converging beam using two identical reverse-oriented prisms that linearly displace each other to perform the correction. On the other hand, RADC employs two counter-rotating amici-prisms to correct both the deviation and dispersion effects simultaneously \cite{Bahrami_11}. However, LADC causes the beam to deviate from its original direction, and this deviation varies with the dispersion correction at different zenith angles. Such beam deviation poses challenges for the fiber input, which must remain stationary to prevent signal loss. Therefore, we decided to incorporate the RADC design into the fiber positioner of the HROS instrument to enhance the precision and efficiency of spectroscopic observations by mitigating the adverse effects of atmospheric dispersion.

\label{section1}

\section{Optical OVERVIEW OF TELESCOPE AND SPECTROGRAPH}
\subsection{TMT and HROS Overview}

TMT is a Ritchy Criterion (RC) telescope that has a hyperbolic primary (M1) and secondary (M2) mirror and a flat tertiary (M3) mirror. The primary mirror has a diameter of 30 meters achieved using 492 hexagonal segments and a conic constant of -1.000953, while the secondary mirror has a diameter of 3.1 meters and a conic constant of -1.318 (see Table \ref{tmttable}). The final focus of the telescope gives an f/15 beam with a plate scale of 2.18mm/1” and a focal plane with a diameter of 3 meters and a radius of curvature of 3.6 meters, which gives a non-telecentric beam. Because of aberrations, TMT's image quality in terms of geometric radius degrades from 20 microns at 0'x0' (on-axis) to 675 microns at 7'x7' (off-axis), as shown in Figure $\ref{TMT_Design}$.

\begin{table}[h]  

    \centering 
    \begin{tabular}{p{8cm}p{4cm}} 
   
    \hline  \
    \centerline {Telescope and Spectrograph Specifications} \\ 
    \hline
     Primary mirror & ${\diameter}$30m \\
     Secondary mirror & $\diameter$3.02m \\
     Radius of curvature of focal plane & $\approx$ 3m\\
     Field of view & 20'(2.56m) \\
     Wavelength coverage & 310 to 1000nm \\
     Red channel range& 310 to 450nm \\
     Blue channel range& 450 to 1000nm \\
     Spectral resolution of MOS & 25,000 \\
     Size of the fiber for MOS & 1" \\
    \hline 
    \end{tabular}
    \caption{TMT and HROS Parameters} 
    \label{tmttable}
\end{table}

The TMT will host adaptive optics and seeing-limited instruments on either side of its Nasmyth platform. HROS is a seeing-limited instrument expected to exploit the TMT's aperture advantage even for a mediocre seeing condition. 
 It will operate in slit and fiber-fed modes and offer high resolution(R=100000), a standard resolution(R=50000), and MOS mode with R=25000.

\section{The optical design of atmospheric dispersion corrector along with its fore optics}
\subsection{ADC Requirement}
\begin{figure}[h!]
    \centering
    \subfigure{\includegraphics[width=1\textwidth]{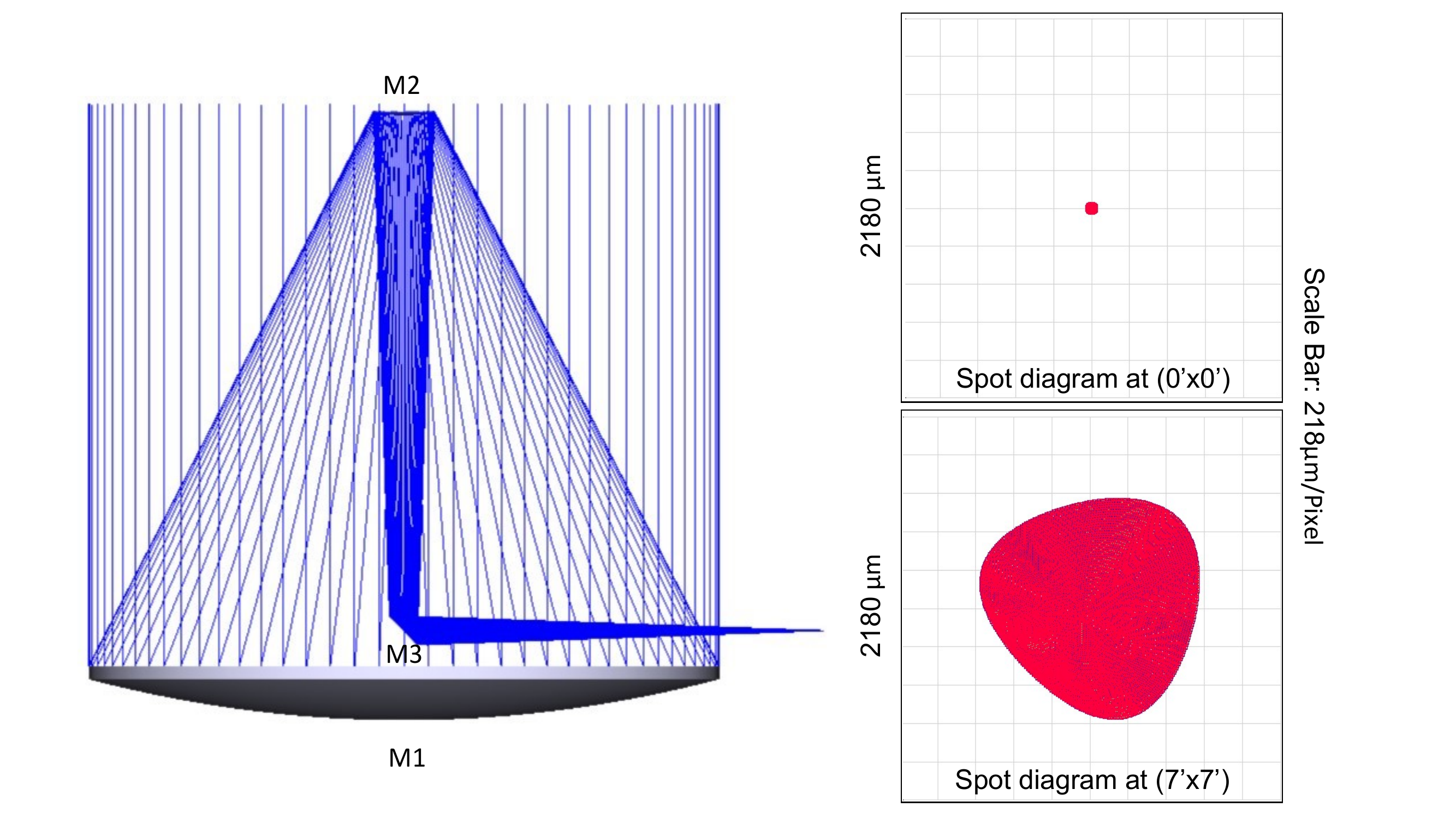}}
    \caption{TMT optical layout and geometric spot diagrams showing the image quality at different fields}
    \label{TMT_Design}

\end{figure}

\begin{figure}[h!]
    \centering
    \subfigure{\includegraphics[trim={0cm 0cm 0.1cm 0cm},clip,width=1\textwidth]{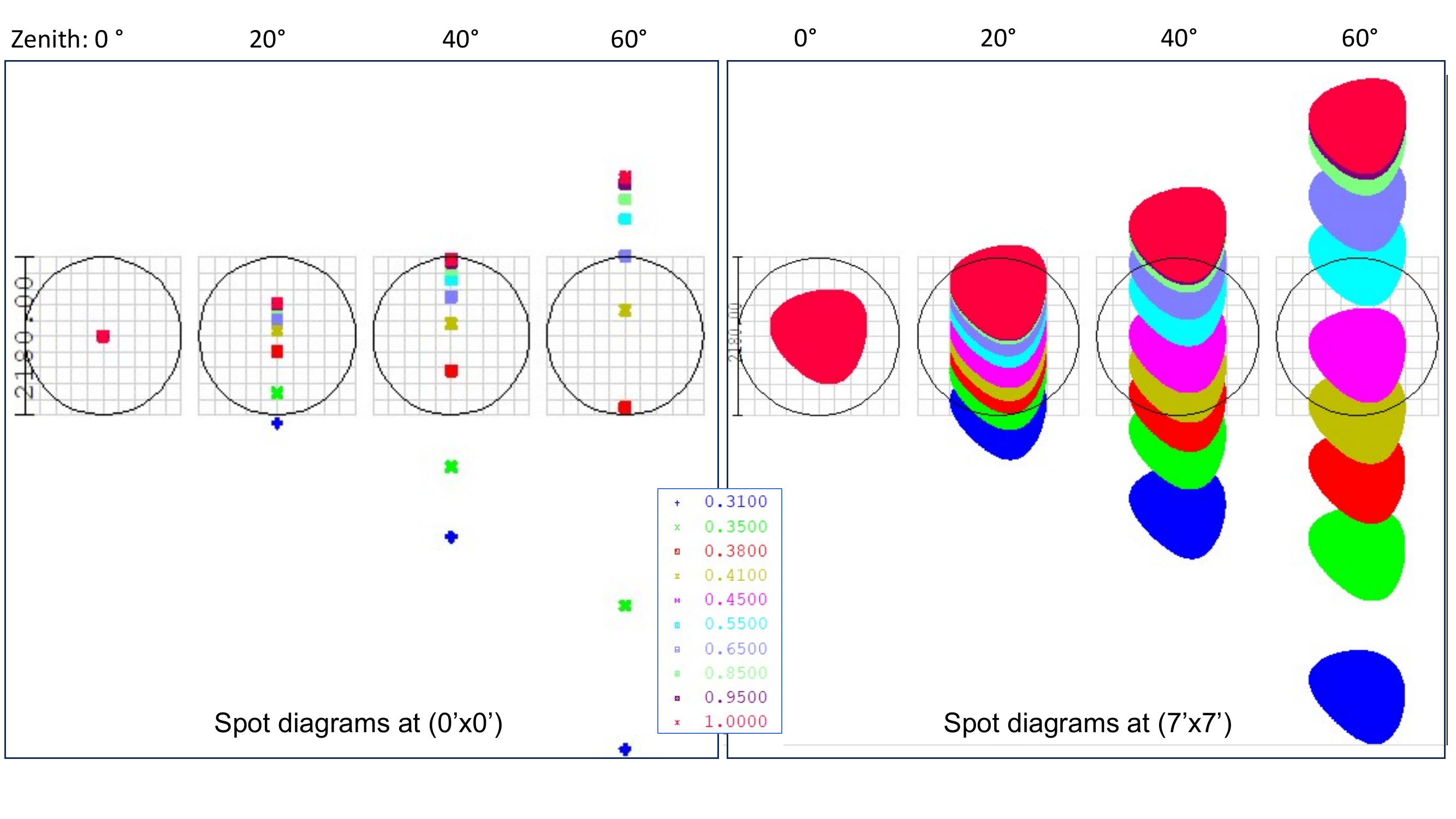}}
    \caption{{ Dispersion uncorrected spot diagrams of two fields at different zenith angles}}
    \label{fig3}
    
\end{figure}

TMT is an alt-azimuth mount telescope that can observe targets from a zenith angle of 0 to 60 degrees. To study the performance of the telescope at various elevations, we generated the telescope model using ZEMAX and obtained the geometric spot diagram at the focal plane as shown in Figure \ref{TMT_Design}. To incorporate the dispersion effect on these spots, we used the atmospheric refraction model\cite{span_14} by inputting the appropriate site conditions (like height, temperature, pressure \& zenith) of Maunakea, Hawaii. The effect of this dispersion on the geometric spots are shown in Figure \ref{fig3}. From these spot diagrams, it is observed that the geometric radius of the image elongates from 20 to about 5000 microns (0.018"- 5") for an on-axis object as it moves to lower elevations.

This will reduce the Geometric Throughput (GT) from about $75\%$ to $45\%$ for the on-axis and $69\%$ to $42\%$ for the off-axis cases (assuming a seeing value of 0.7") for input to a 1" circular fiber over a range of wavelengths from 310 to 1000nm. The methodology for measuring GT is described in Section \ref{section4}. To achieve about $70\%$ throughput even at lower elevations, we need an ADC, that can correct the dispersion with minimal deviation. The optical design of the ADC is discussed in the following section.

\vspace{500pt}

\subsection{ADC Optical Design}

\begin{figure}[h]
    \centering
   \includegraphics[width=18cm]{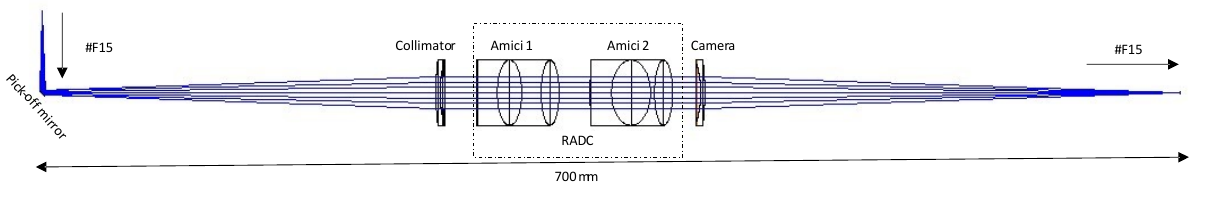} 
    \caption{Optical Layout of ADC at Zenith 0\degree}
    \label{fig:TMT}
\end{figure}

The optical layout of the proposed instrument is given in Figure \ref{fig:TMT}. The design includes a pick-off mirror of about 5" in size that can tilt to direct the non-telecentric f/15 beam from the TMT focal plane (either on-axis or off-axis beam) to the collimator. The pick-off mirror has to make a $45\degree$ angle with the tangent plane of the image chief ray to direct the beam towards the collimator. Hence, the angle of the pick-off mirror with the chief ray at various field positions has to be equal to $45\degree$-the angle between the tangent plane of an on-axis object and the tangent plane of the off-axis object. The collimator is made of a doublet lens with an effective focal length of 291mm and materials Nikon 7054 and S-FPL51Y, which has more than $95\%$ transmission (per 5mm thickness) over 310 to 1000nm. The collimated beam then passes through the ADC.

\begin{table}[h!]  
    \caption{Specifications of ADC  and  fore optics} 
    \centering 
    \begin{tabular}{p{8cm}p{5cm}} 
    \hline  \
    \centerline {Atmospheric Dispersion Corrector} \\ [0.5ex]   
    \hline
     Diameter & 30mm \\
     Materials & Nikon 7054\&CAF2 \\
     Apex Angles of \\ P1,P2,P3 and P4 & 22.492\degree,37.696\degree,37.289\degree,21.492\degree \\
     Counter rotation angles of amici-prisms\\ at zenith angles 0\degree,20\degree,40\degree,60\degree  & 90\degree,78.95\degree,61.705\degree,2.895\degree \\
    \hline  \
    \centerline {Collimator and Camera} \\ [0.5ex]  
    \hline
     Diameter & 30mm \\
     Effective focal length & 291mm \\
     Materials &  S-FPL51Y\&7054\\ 
    \hline 
    \end{tabular}
    \label{table2}
\end{table}

\begin{figure}[h!]
    \centering
    \subfigure{\includegraphics[width=1\textwidth]{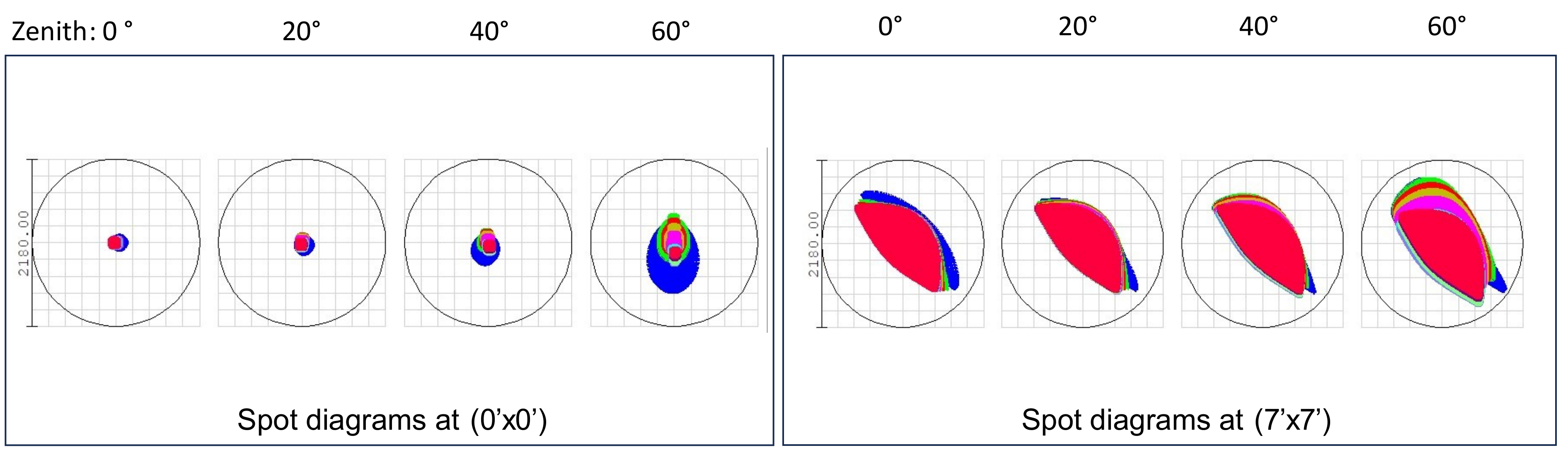}}

    \caption{Dispersion corrected spot diagrams of two fields at different zeniths}
    \label{fig5}
\end{figure}

As mentioned in Section \ref{section1}, we use RADC instead of LADC to correct both dispersion and deviation. The RADC consists of two doublet prisms called amici-prisms comprised of four circular prisms (P1, P2 \& P3, P4, respectively) of Nikon 7054 and CAF2 materials. The apex angles of the four individual prisms are chosen such that the second prism(P2, P4) corrects the deviation caused by the first prism (P1, P3)  of both amici prisms, and both amici-prisms counter-rotate to cancel out the dispersion. To get these optimal apex angles and counter-rotation angles of amici prisms, we utilized the multi-configuration editor and merit function operands of ZEMAX. The optimized values are given in Table \ref{table2}. A camera (with the same specifications as the collimator) has been used to focus the dispersion-corrected beam coming from ADC.
The geometric spot diagrams after the dispersion correction are shown in Figure \ref{fig5}.

\section{Instrument Performance Estimates}

\label{section4}

\subsection{Throughput}

\begin{figure}[h]
  
    \centering
    \subfigure[Image at Zenith 0\degree]{\includegraphics[width=0.3\textwidth]{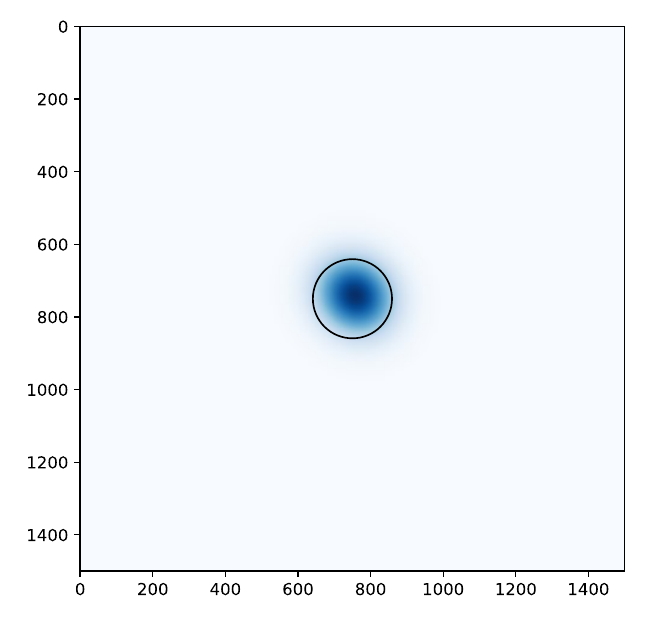}}
    \hfill
    \subfigure[Dispersed Image at Zenith 60\degree]{\includegraphics[width=0.3\textwidth]{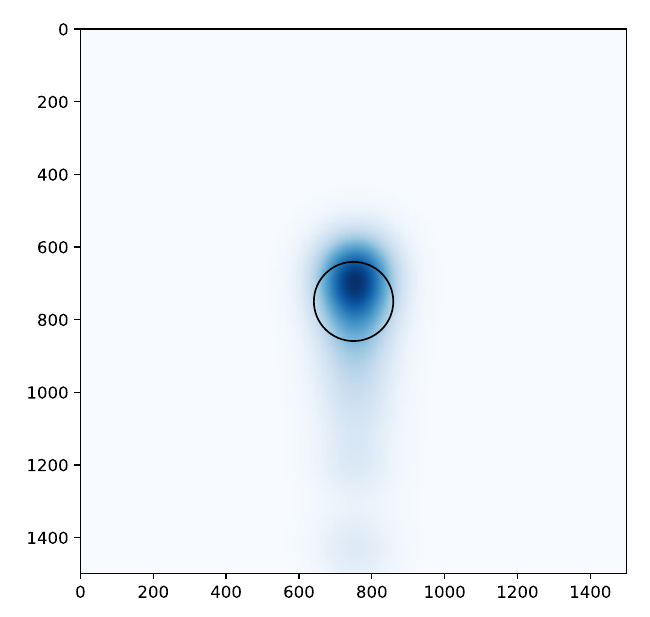}}
    \hfill
    \subfigure[Dispersion Corrected Image at Zenith 60\degree]{\includegraphics[width=0.28\textwidth]{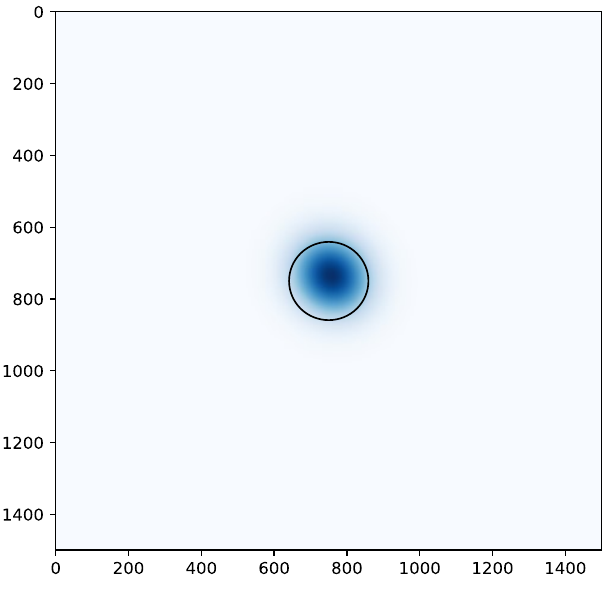}}\\
              Scale bar:100units = 1mm\\
    \caption{Python-ZEMAX simulated images of seeing included geometric spot diagrams at 
    7'x7'}
    \label{seeing_image}

\end{figure}

We used the geometrical image analysis (IMA) tool in  ZEMAX and the PyZDDE\cite{indranil_2014} application program interface(API) to estimate the geometric throughput of the on-axis object for 1" fiber with and without dispersion correction. The methodology for measuring geometrical throughput involved calculating the seeing (2d Gaussian profile) for each wavelength, using the Fried parameter (r0) of 0.7" at 550nm. This was convolved with the extracted geometrical position-based intensities obtained from IMA and generated the images of seeing included geometric spots for different zenith angles before and after dispersion correction. The resultant geometric spots for 0\degree and 60\degree zenith angles are shown in Figure \ref{seeing_image}. Finally, the result was multiplied by a 1" python simulated circular aperture (representing the 1" fiber). The percentage of geometric throughput($\%GT$) is then measured using Equation \ref{GP}. 

\begin{gather}
    \text{\%GT =} {\text{(OPI/IPI)*100}}\label{GP}
\end{gather}

\noindent IPI = Sum of intensity before multiplication with aperture(Input Intensity)\\
OPI = Sum of intensity after multiplication with aperture(Output Intensity)

\begin{figure}[h]
    \centering
    \subfigure[Before dispersion correction]{\includegraphics[width=0.45\textwidth]{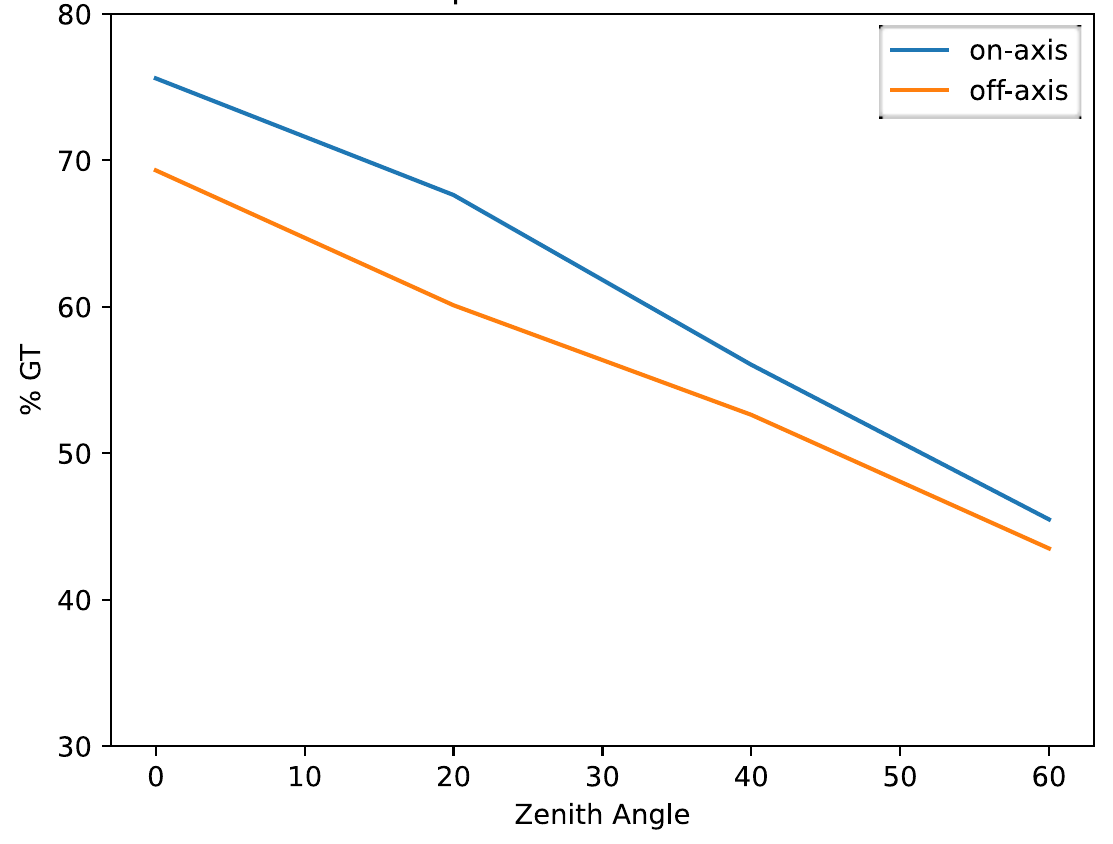}}
    \subfigure[After dispersion correction]{\includegraphics[width=0.45\textwidth]{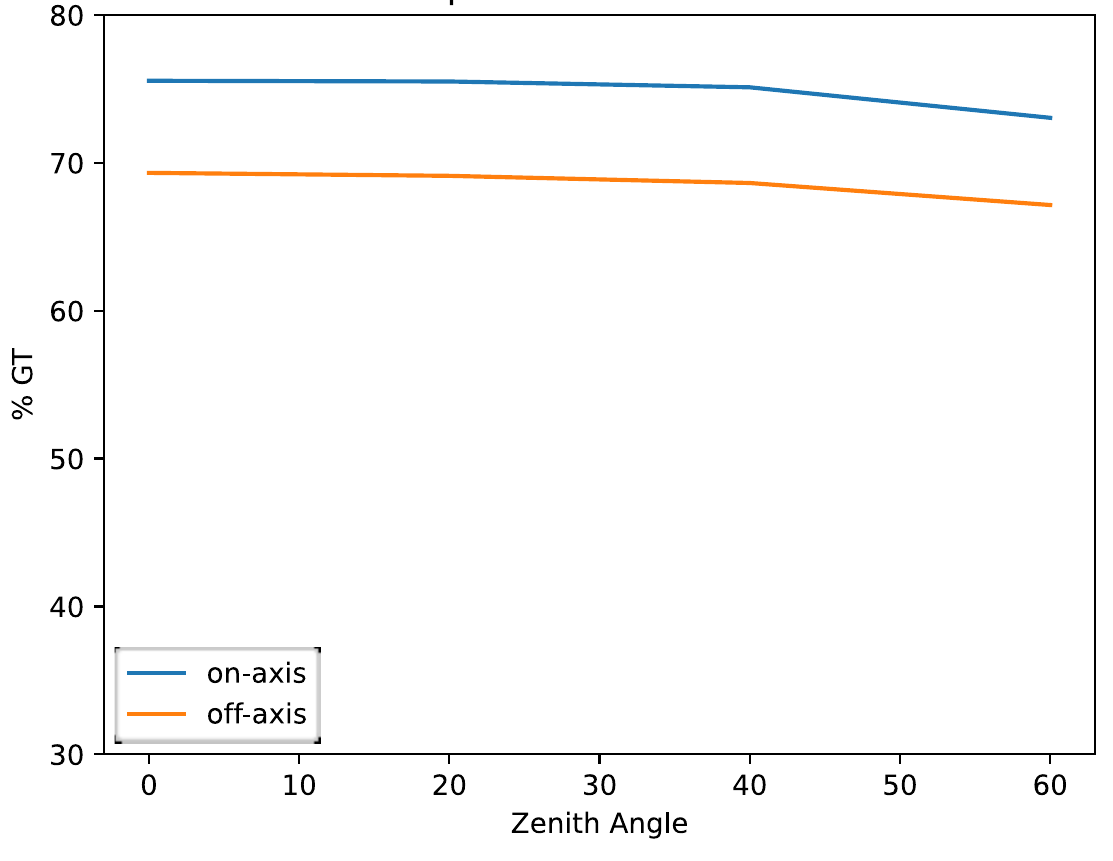}}
    \caption{Geometric throughput for 1" circular fiber }
    \label{figure6}
\end{figure}

\noindent The obtained percentage of geometrical throughput for different zenith angles for an on-axis object before and after dispersion correction are shown in Figure \ref{figure6}.

\subsection{SNR}

\begin{figure}[h]
    \centering
    \subfigure[Before dispersion correction]{\includegraphics[trim={0.0cm 0.0cm 0.0cm 0.0cm},clip,width=0.45\textwidth]{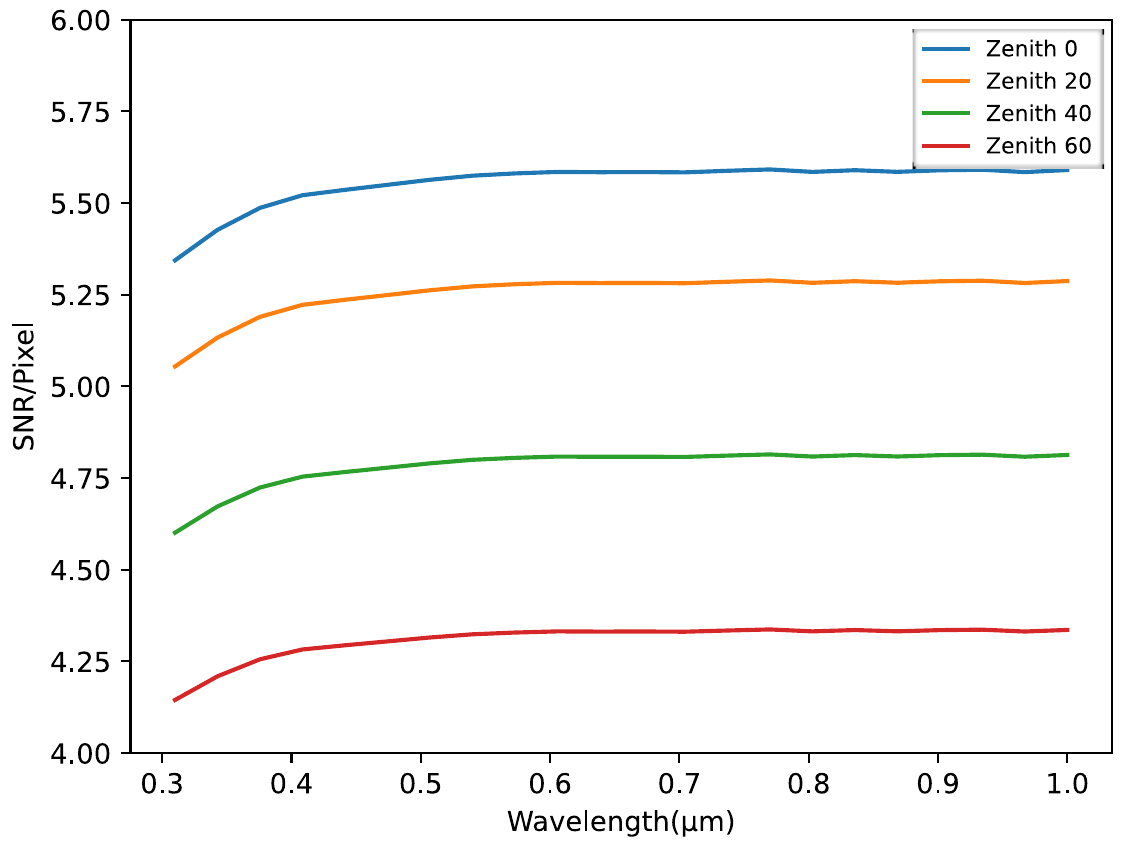}}
    \subfigure[After dispersion correction]{\includegraphics[trim={0.0cm 0.0cm 0.0cm 0.0cm},clip,width=0.45\textwidth]{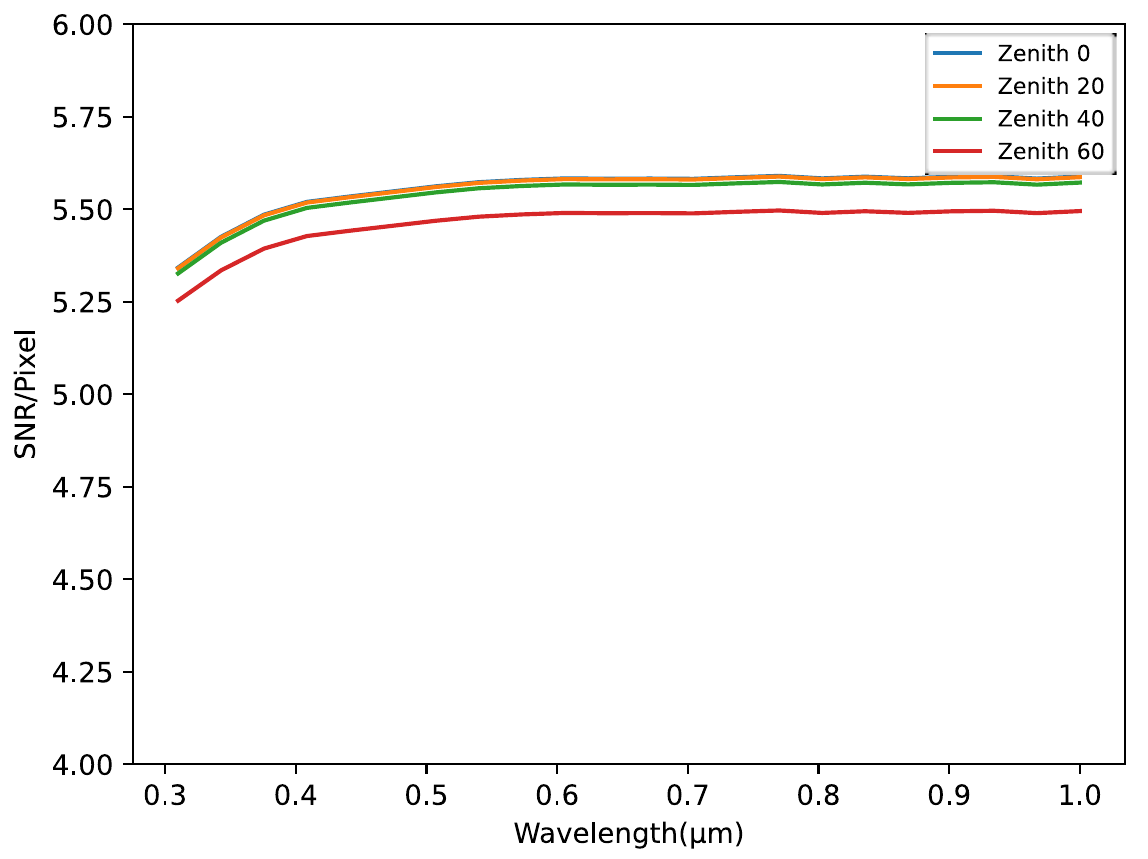}}
    \caption{Wavelength vs. SNR/pixel before and after dispersion correction of an on-axis object}
    \label{figure7}
\end{figure}

The Signal-to-Noise Ratio (SNR) per pixel was estimated before and after dispersion correction through a series of calculations. Initially, the number of photons per cm\textsuperscript{2} per second (N) from an object with a magnitude of 15 was considered, taking into account a spectrograph resolution (R) of 25000 for MOS and accounting for the Point Spread Function (PSF) width of 3.5 pixels. This value is multiplied by the total system response (expressed in area units), as described in \cite{Bharat_2020}. The total system response is determined by considering various factors, including the collecting area (A), the transmissivity of the lenses and prisms (t), the reflectivity of mirrors (r) within the system, as well as the spectrograph efficiency ($\eta\approx$ 10$\%$). Moreover, we considered the transmission of fiber per meter ($t_f$) along with the percentage of GT, and the  detector's Quantum Efficiency (QE$\approx$ 90$\%$) for the calculation. Equation $\ref{snr}$ illustrates the final computation of the SNR per pixel.

\begin{gather}
    \text{Total System Response (SR) =} A * r(\lambda) * t(\lambda) * t_f(\lambda) * \%GT(zenith) * \eta * QE.  
\end{gather}

\begin{gather}
    \text{SNR/Pixel =} \sqrt{(N/(R*PSF width))*SR}\label{snr}
\end{gather}
\noindent The resultant SNR per pixel values for different wavelengths and zenith angles before and after dispersion correction are shown in Figure \ref{figure7}.

\subsection{Tolerance analysis}

We used the tolerancing tool of ZEMAX to perform tolerance analysis for the optical design. This helped us evaluate how manufacturing and alignment errors affect the system's performance.  The input values for the possible errors are decided based on the manufacturer tolerances for this design. We observed that the average image quality would decrease by 5$\%$ for the given tolerances as shown in Table $\ref{tolarence}$.

\begin{table}[ht]
\centering
\begin{tabular}{|c|c|c|c|}
\hline
Collimator\& Camera Parameters & Range & ADC Parameters & Range\\
\hline
Thickness & $\pm$ 0.1mm & Apex Angle & 0.5$\degree$\\
Radius & $\pm$1\% & X Decenter & $\pm$2.5mm\\
X-Tilt & $\pm$1' & Y Decenter & $\pm$2.5mm\\
Y-Tilt & $\pm$1' & & \\
X-Decenter & $\pm$1mm & & \\
Y-Decenter & $\pm$1mm & & \\
\hline
\end{tabular}
\caption{Tolerance values for Collimator\&Camera and ADC}
\label{tolarence}
\end{table}

\section{Conclusion and Future Work}

The proposed ADC design shows improved SNR at higher zenith angles by correcting broad wavelength coverage spanning 310-1000nm, as shown in Figure \ref{figure7}. Additionally, the range of deviations after the correction is confined within a 0.125" radius. The tolerance analysis confirms that the image degradation is merely 5\%, thereby having a minimal effect on SNR. To further enhance the design, which operates in two distinct wavelength ranges (as outlined in Section \ref{sec:intro}), a dichroic will be introduced behind the camera to split the beam into separate channels efficiently. Incorporating microlens into the design will allow a transition from f/15 to f/3.1, enabling effective fiber coupling. Furthermore, to validate the simulated outcomes, we will compare them with the observational results obtained from existing telescopes.

\acknowledgments 

Manjunath would like to acknowledge Radhika Dharamadhikari's help with ZEMAX, Saraswathi Kalyani's suggestions on calculations, and Bernard Delabre for suggesting the Nikon 7054 glass material. We also acknowledge the Department of Science and Technology and the Indian Institute of Astrophysics for their support during this work.
 
\bibliography{report} 

\bibliographystyle{spiebib} 
\end{document}